\pdfoutput=1
\documentclass[pdftex,preprint,12pt]{elsarticle}

\newcommand{\lsim}{\raisebox{-0.13cm}{~\shortstack{$<$ \\[-0.07cm] $\sim$}}~} 
\newcommand{\gsim}{\raisebox{-0.13cm}{~\shortstack{$>$ \\[-0.07cm] $\sim$}}~}

\newcommand{\tb}{\tan\beta} 
\newcommand{\beq}{\begin{eqnarray}} 
\newcommand{\eeq}{\end{eqnarray}}

\usepackage{graphicx}
\usepackage{amssymb}
\journal{Physics Letters B}

\begin{document}

\begin{frontmatter}

%% Title, authors and addresses

%% use the tnoteref command within \title for footnotes;
%% use the tnotetext command for the associated footnote;
%% use the fnref command within \author or \address for footnotes;
%% use the fntext command for the associated footnote;
%% use the corref command within \author for corresponding author footnotes;
%% use the cortext command for the associated footnote;
%% use the ead command for the email address,
%% and the form \ead[url] for the home page:
%%
%% \title{Title\tnoteref{label1}}
%% \tnotetext[label1]{}
%% \author{Name\corref{cor1}\fnref{label2}}
%% \ead{email address}
%% \ead[url]{home page}
%% \fntext[label2]{}
%% \cortext[cor1]{}
%% \address{Address\fnref{label3}}
%% \fntext[label3]{}

\title{An update of the constraints on the phenomenological MSSM from the new LHC Higgs results}

%% use optional labels to link authors explicitly to addresses:
%% \author[label1,label2]{<author name>}
%% \address[label1]{<address>}
%% \address[label2]{<address>}

\author[Lyon1,Lyon2,CERN]{Alexandre~Arbey}
\author[UCSC,CERN]{Marco~Battaglia}
\author[ORSAY]{Abdelhak~Djouadi}
\author[CF,CERN]{Farvah~Mahmoudi}
\address[Lyon1]{Universit\'e de Lyon, France; Universit\'e Lyon 1, F-69622~Villeurbanne
Cedex, France}
\address[Lyon2]{Centre de Recherche Astrophysique de Lyon, Observatoire de Lyon, Saint-Genis Laval Cedex, 
F-69561, France; CNRS, UMR 5574; Ecole Normale Sup\'erieure de Lyon, Lyon, France}
\address[CERN]{CERN, CH-1211 Geneva 23, Switzerland}
\address[UCSC]{Santa Cruz Institute of Particle Physics, University of California, Santa Cruz,
CA 95064, USA}
\address[ORSAY]{Laboratoire de Physique Th\'eorique, Universit\'e Paris XI and CNRS,
F-91405 Orsay, France}
\address[CF]{Clermont Universit\'e, Universit\'e Blaise Pascal, CNRS/IN2P3, LPC, BP 10448, 
F-63000 Clermont-Ferrand, France}

\begin{abstract}
Updated results on the search of Higgs bosons at the LHC with up to 17~fb$^{-1}$
of data  have just been presented by the ATLAS and CMS collaborations. New
constraints are provided  by the LHCb and XENON experiments with the observation
of the rare decay $B_s \to \mu^+ \mu^-$   and new limits  on dark matter direct
detection. In this paper, we update and extend the results on  the implications
of these data on the phenomenological  Minimal  Supersymmetric extension of  the
Standard Model (pMSSM) by using high statistics, flat scans of its 19
parameters. The new LHC data on $b\bar b$ and $\tau\tau$ decays of the lightest
Higgs state and the new CMS limits from the $\tau\tau$ searches for the heavier
Higgs states set stronger constraints on the pMSSM parameter space. 
\end{abstract}

\begin{keyword}
%% keywords here, in the form: keyword \sep keyword

%% MSC codes here, in the form: \MSC code \sep code
%% or \MSC[2008] code \sep code (2000 is the default)

\end{keyword}

\end{frontmatter}

%\linenumbers

\section{Introduction}
\label{sec1}

The first results on the mass and decay rates of the Higgs-like particle
observed by ATLAS and CMS at the  LHC \cite{ATLAS:2012zz,CMS:2012zz} already
imply some significant bounds on the parameters of the Minimal Supersymmetric
extension of the Standard Model  (MSSM), once we interpret the newly discovered
particle as the lightest $h$ state in this theory. New results for the 
properties of this particle have just been presented by the ATLAS and CMS
experiments with up to 17~fb$^{-1}$ of 7 and  8~TeV data
\cite{ATLAS-2012-168,ATLAS-2012-169,CMS-12-015,CMS-12-016,ATLAS-2012-158,ATLAS-2012-161,
CMS-12-044,ATLAS-2012-160,CMS-12-043}. New results on the search for tau lepton
pairs at high invariant masses reported by the CMS collaboration \cite{CMS-12-050}
set tighter constraints on heavier  Higgs particles. On another front the LHCb
collaboration has obtained the first observation of the $B^0_s \to \mu^+ \mu^-$ 
rare decay and reported a first determination of its decay branching fraction
\cite{Aaij:2012hcp}. Constraints on weakly interacting dark matter particles
have also been significantly improved since the Higgs-like particle discovery
with the updated results on the  direct search by the XENON collaboration
\cite{Aprile:2012nq}. All these new data have a drastic impact not only on the
SM but  also on theories of supersymmetry and in particular the MSSM.

In Ref.~\cite{Arbey:2012dq} we presented a detailed analysis of the implications
of the observation of a Higgs-like  particle and the first determination of its
properties. There, we refined a previous study \cite{Arbey:2011ab} of the
implications of the value $M_h \approx 126$ GeV for   both the constrained and
unconstrained versions of the MSSM and analysed the impact of the first data
for the newly observed particle. By reviewing the different regimes of the
MSSM, we concluded that two of those exhibited the best agreement with
its properties: i) the decoupling regime in which the $h$ boson has
SM-like properties with the  $H$, $H^{\pm}$ and $A$ bosons being heavy and
decoupled from the gauge bosons and ii) a regime where light SUSY particles such
as tau--sleptons, charginos  and third generation scalar quarks affected the $h$
boson rates, in particular in the $h\to \gamma \gamma$ channel.

The study presented in this paper updates that work with a focus on the regions
of pMSSM parameters allowed, and favoured,  by the latest LHC Higgs data and
other results. We evaluate the constraints obtained in the framework of
the phenomenological  MSSM (pMSSM), with the neutralino as the lightest SUSY
particle (LSP), with 19 free parameters using flat parameter scans.  Our
analysis tests the compatibility of a large sample of generated pMSSM points,
fulfilling the constraints from other MSSM  searches at LEP and the LHC, flavour
physics data and dark matter searches and the direct searches for supersymmetric
particles in channels with missing  transverse energy  at the LHC. The
qualitative results of our previous study~\cite{Arbey:2012dq} stay the same, and
are even strengthened; there are quantitative changes which make  this
update interesting. While the statistical accuracy of the LHC results is  still
limited and the data have not settled, we expect a steady improvement with the
analysis of the  full statistics collected in the 8~TeV run and
then with higher energy LHC operation from 2014.  The present study provides
conclusions which are already a useful guidance for the current SUSY searches at
the LHC and  it defines a template for further analyses, once results with
better precision will be available. 

The essential elements of the pMSSM scans with the various 
constraints and the relevant ranges for the variation of its parameters have been 
already  presented in Ref.~\cite{Arbey:2012dq}. 
The tools used to perform our analysis given in Ref.~\cite{tools}. Here
we proceed to the presentation of our updated
analysis and its results in the next section. Section~3 has a short conclusion.

\section{Analysis and Results}
\label{sec3}

The analysis is based on the latest results for the mass of the new  Higgs--like
particle and its signal strengths in the individual  channels. We use a weighted
average of the results reported by the ATLAS and CMS collaborations at the LHC
and also CDF and D0 at the Tevatron \cite{Tevatron:2012zz}  with their estimated
statistical uncertainties, as summarised in Table~\ref{tab:input}.  In the
following, we use the notation $R_{XX}$ to indicate the ratio of the $h$
branching fraction  to the final state $XX$, BR($h \rightarrow XX$), to  its SM
value.  Then, we compute the so-called ``signal strengths'', {\it i.e.} the
ratios of the products of production cross sections  times decay branching
fractions for the pMSSM points to their SM values, which we denote with
$\mu_{XX}$ for a given  $h\to XX$ channel, $\mu_{XX} = \sigma (h)/ \sigma
(H_{\rm SM}) \times R_{XX}$, where $\sigma$ is the relevant production  cross
section. 

\begin{figure}[ht!]
\vspace*{-3mm}
\begin{center}
\hspace*{-0.50cm} \includegraphics[width=6.5cm]{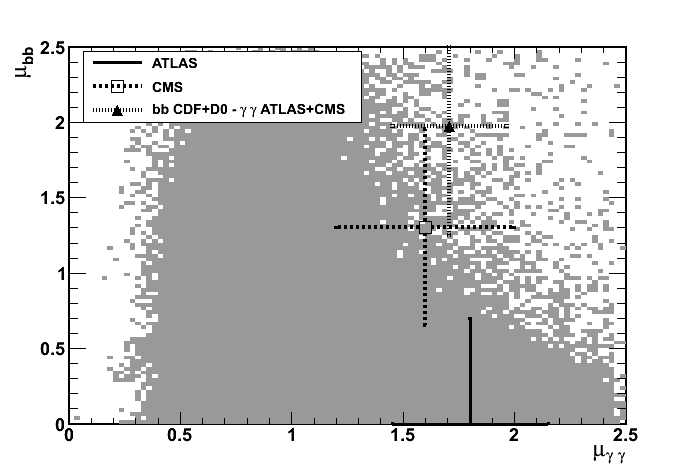} \\
\hspace*{-0.50cm} \includegraphics[width=6.5cm]{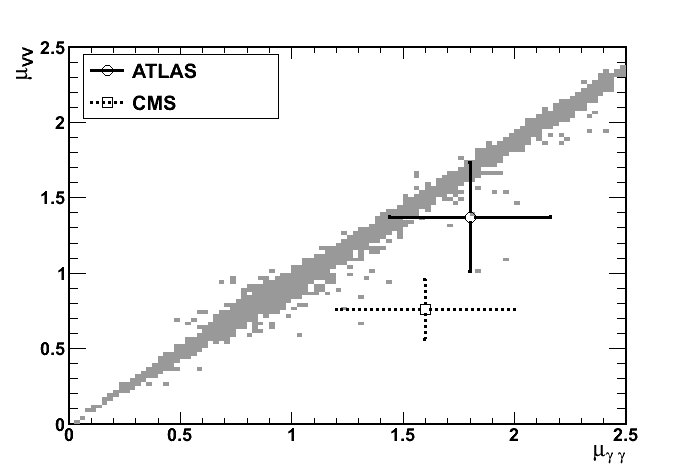} \\
\hspace*{-0.50cm} \includegraphics[width=6.5cm]{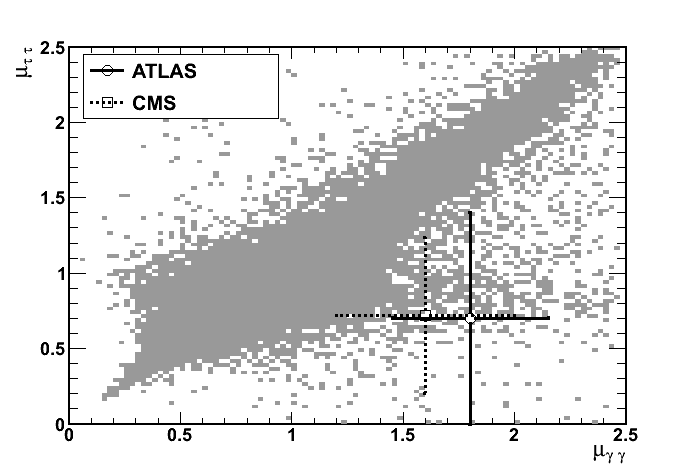} \\
\end{center}
\vspace*{-5mm}
\caption{\small The $b\bar b$ (top), $VV\!=\!WW\!+\!ZZ$ (centre) and $\tau \tau$ (bottom) 
signal strengths vs.\ that for $\gamma \gamma$  from the LHC and Tevatron results compared to the values for the accepted 
pMSSM points.}
\label{fig:Mu}
\end{figure}

The signal strengths corresponding to each accepted pMSSM point are compared to their experimental values. 
Both ATLAS and CMS have provided updates for the $ZZ$, $WW$, $b\bar b$ and $\tau \tau$ channels with the 
full 7~TeV statistics of 4.7~fb$^{-1}$ and $\simeq$~13~fb$^{-1}$ of the 8~TeV data. ATLAS has also updated the 
result for the $\gamma \gamma$ channel based on 4.8 + 13~fb$^{-1}$. 
These updates result in improved determinations of the signal strengths in the $\gamma \gamma$, $WW$ and $ZZ$ channel and data 
from both LHC experiments in the important $b\bar b$ and $\tau \tau$ channels. We also include the combined Tevatron result 
for the $b \bar b$ channel. While these results are compatible with the SM expectations, or the MSSM in the decoupling 
regime with heavy SUSY particles, within their present accuracy, the situation 
with the data for most of the channels does not appear to have settled. After the ATLAS update, the results for the 
$\gamma \gamma$ channel hint more significantly to a possible enhancement of its rate, but a confirmation from CMS with the 
larger 8~TeV data set would be comforting. Results on the 
$WW$ and $ZZ$ yields are aligned at values which are consistently about 1$\sigma$ above the SM expectation for ATLAS and 
1$\sigma$ below it for CMS. The important, but experimentally difficult, $b\bar b$ channel still requires more data and a 
careful control of the SM backgrounds. At present, the spread of the experimental results by ATLAS and CMS, covers the 
range of values predicted by the MSSM, as we highlight in Figure~\ref{fig:Mu}.

\begin{table}[!h]
\begin{center}
\begin{tabular}{|c|c|c|}
\hline
Parameter & Value & Experiment \\ \hline \hline
$M_h$     & 126$\pm$2 GeV & ATLAS \cite{ATLAS:2012zz} + CMS \cite{CMS:2012zz} \\ 
$\mu_{\gamma \gamma}$ & 1.71$\pm$0.26 & ATLAS \cite{ATLAS-2012-168} + CMS \cite{CMS-12-015} \\
$\mu_{Z Z}$ & 0.97$\pm$0.26 & ATLAS \cite{ATLAS-2012-169} + CMS \cite{CMS-12-016} \\
$\mu_{W W}$ & 0.85$\pm$0.23 & ATLAS \cite{ATLAS-2012-158} + CMS \cite{CMS-12-016} \\                                 
\hline
$\mu_{b \bar b}$  & 1.28 $\pm$ 0.45 & ATLAS \cite{ATLAS-2012-161} + CMS \cite{CMS-12-044} + (CDF + D0) \cite{Tevatron:2012zz}\\ 
$\mu_{\tau \tau}$ & 0.71 $\pm$ 0.42 & ATLAS \cite{ATLAS-2012-160} + CMS \cite{CMS-12-043}\\ 
\hline
$D_{\gamma \gamma}$ &  1.88$\pm$0.46 &                                           \\
$D_{\tau \tau}$     &  0.79$\pm$0.49 &   \\    
\hline
\end{tabular}
\end{center}
\caption{\small Input values for the Higgs mass and rates used for the study.}
\label{tab:input} 
\end{table}

We perform an analysis of the compatibility of the MSSM with these results, 
based on our set of $2.0 \times 10^8$ pMSSM points, with the assumption that
the  observed particle is the lightest Higgs boson of the MSSM, $h$, and we
comment on the possible identification  of the particle with the $H$ state later
in this section. We first select points fulfilling  all the constraints
discussed  in Ref.~\cite{Arbey:2012dq}  but including the new measurements. In
particular, for the decay $B_s \to \mu^+ \mu^-$, the LHCb collaboration 
reported a branching ratio of BR($B_s^0 \to \mu^+ \mu^-) = (3.2^{+1.5}_{
-1.2})\times 10^{-9}$  with 2.2~fb$^{-1}$ of data~\cite{Aaij:2012hcp}. We derive the 
constraint at 90\% C.L. after accounting for theoretical uncertainties, estimated to be 
at the  11\% level~\cite{Mahmoudi:2012un}. For the
$b\to s\gamma$ decay branching ratio, the new world average value  of $(3.43 \pm
0.22)\times 10^{-4}$ \cite{bsgamma} is now closer to the SM prediction 
leading to more severe constraints.  The dark matter constraints are also
updated for the  direct  detection limits by including the new result on the
spin-independent  $\chi$-nucleon scattering cross section from 225 live days of
XENON-100 data~\cite{Aprile:2012nq}. We note that for the Higgs decay branching
fractions, we use the latest version of {\tt HDECAY (5.0)}~\cite{Djouadi:1997yw}
which includes, among the new features,  a more refined treatment of the SUSY
vertex corrections.

Then, we move to consider the compatibility of the $h$ mass  and signal
rates predicted for the accepted pMSSM points with the updated LHC
measurements. The results on the signal strengths from the current results for the individual 
experiments are compared to the distributions obtained for accepted pMSSM points in 
Figure~\ref{fig:Mu}. We use the signal strengths for the channels
where a signal  has been observed, $\mu_{\gamma \gamma}$, $\mu_{Z Z}$,  $\mu_{W
W}$ and we also add the limits obtained for $\mu_{b \bar b}$ and $\mu_{\tau
\tau}$.  The systematic uncertainties from the Higgs production cross section in
the $gg \to h$ channels may be  sizeable~\cite{THU}, at least $\pm$15~\%, and
have been taken into  account. The use of ratios
of the signal strengths reduces these uncertainties and we thus
also test the signal strength ratios $D_{\gamma \gamma}$ =
$\mu_{\gamma \gamma} / \mu_{VV}$ and  $D_{\tau \tau}$ = $\mu_{\tau \tau} /
\mu_{VV}$ \cite{Djouadi:2012he}, where $\mu_{VV}$  is the weighted average of the signal 
strengths in the $WW$
and $ZZ$ channels, which are mostly immune from these  systematics

In order to evaluate the compatibility of each point with the Higgs results, we compute the total 
$\chi^2$ probability for the observables of Table~\ref{tab:input} for each 
accepted pMSSM point. 
The $\chi^2$ for a given pMSSM point is built as 
\begin{eqnarray}
\chi^2 = \frac{(M_{h}(LHC) - M_{h}(i))^2}{\delta^2 [M_{h}(LHC)] + \delta^2 [M_{h}(th)]} + 
\sum_j \frac{(\mu_{j} (LHC) - \mu_{j} (i))^2}{\delta^2 [\mu_{j}(LHC)] + \delta^2 [\mu_{j}(th)]}
\end{eqnarray}
where $i$ is the index of the pMSSM point, $M_{h}(LHC) \pm \delta [M_{h}(LHC)]$ and
$\mu_{j} (LHC) \pm \delta [\mu_{j}(LHC)]$ the LHC (and Tevatron) 
measurements of the mass and the signal strengths in channel $j$ 
with their uncertainties as given in Table~\ref{tab:input} and the theory uncertainties $\delta^2 [M_{h}(th)]$ 
and $\delta^2 [\mu_{j}(th)]$ account for the theory systematics on the MSSM 
$h$ mass, $\pm$1.5~GeV and the 
production rate.
 
For the $b \bar b$ and $\tau^+ \tau^-$ channels, where no signal evidence has been reported, we add the 
contribution to the total $\chi^2$ only when the respective $\mu$ value is outside the $\pm$1.5~$\sigma$ 
interval from the measured central value, and the pMSSM point becomes increasingly less consistent with the 
limits reported by the LHC and Tevatron experiments.

\subsection{SUSY corrections to the Higgs rates}

\begin{figure}[ht!]
\vspace*{-3mm}
\begin{center}
\includegraphics[width=8.0cm]{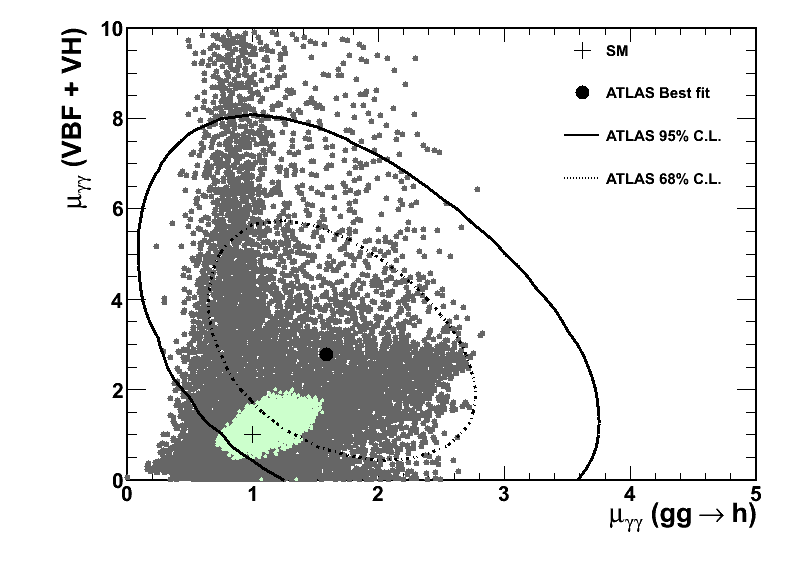} \\
\end{center}
\vspace*{-10mm}
\caption{\small $\mu$ values in the $h \to \gamma \gamma$ channel for associate VBF and VH production 
vs.\ $gg \to h$. The dots in dark grey show the accepted pMSSM points and those in light green the points 
which agree at 90\% C.L. with the constraints of Table~1. The contours give the results obtained by the ATLAS 
experiment (adapted from Ref.~\cite{ATLAS:2012zz}).}
\label{fig:VBF}
\vspace*{-2mm}
\end{figure}

In general, deviations of the $\mu$ signal strength ratios from their SM values
may be due to modifications of either  the decay branching
fractions or the relevant production cross sections,  or
to both. In order to disentangle these effects, it is important to conduct
analyses where the same decay channel  is studied in different production
processes, such as gluon fusion $gg\to h$, associated production with a
gauge boson (VH)  or forward jets (VBF).  The ATLAS collaboration published a
first attempt to separate the contribution of the VBF and VH  production from
$gg \to h$ in the $h \to \gamma \gamma$ channel~\cite{ATLAS:2012zz}.  The
confidence level (C.L.) contours obtained in the analysis are compared in
Figure~\ref{fig:VBF} to the  distribution for all the accepted pMSSM points and
to those selected within the 90\% C.L. with the  Higgs results from the
$\chi^2$ probability analysis.

\begin{figure}[ht!]
\vspace*{-4mm}
\begin{center}
\begin{tabular}{cc}
\hspace*{-0.60cm} \includegraphics[width=7.5cm]{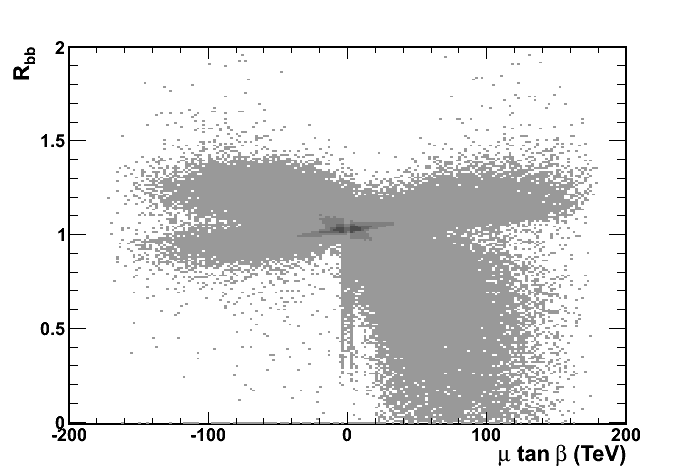} &
\hspace*{-0.80cm} \includegraphics[width=7.5cm]{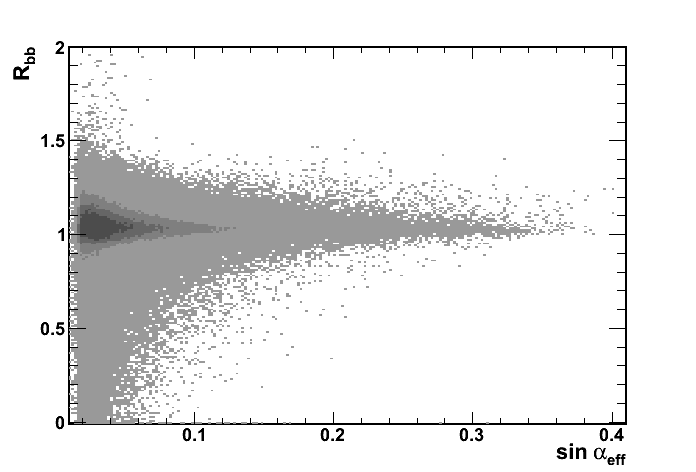} \\
\end{tabular}
\end{center}
\vspace*{-6mm}
\caption{\small Dependence of the $h \to b \bar b$ branching fraction normalised to the SM 
expectation as a function of $\mu \tan \beta$ (left) and  $\sin \alpha_{\mathrm{eff}}$ 
(right). The intensity of the grey tones is proportional to the density of pMSSM points, which 
are peaked around $R_{bb} \sim 1$ }
\label{fig:Rbb}
\vspace*{-1mm}
\end{figure}

The $h$ decay branching fractions may be modified by a change of the $h$ total
decay width. Since the dominant decay mode for a $\sim$126~GeV lightest $h$ boson is $
h \to b \bar b$, a change of the effective $hb\bar b$ coupling by direct vertex
corrections, through the $\Delta_b$ correction that grows as $\mu \tan \beta$, 
results in an anti-correlated variation of the branching fractions of all the
other modes compared to that in $b \bar b$.  The reduction of the $h\to b\bar b$
decay width, away from the decoupling regime  $M_A \gg M_Z$, occurs in a
non-trivial way. The radiative corrections  to the mixing angle $\alpha$ in the 
CP--even Higgs sector strongly affect the $hb\bar b$ coupling, $g_{hb \bar b}=
-\sin\alpha_{\rm eff}/\cos\beta$. While in the decoupling limit we expect 
$\tan \alpha_{\rm eff}  \to -1/\tan \beta$ making $g_{hb \bar b}$  to
become SM--like, there is a  combination of parameters which realises the
so--called ``vanishing coupling'' regime \cite{vanishing} in which $\alpha_{\rm eff} \to 0$.
In this case, ($\tan \alpha_{\rm eff} \tan \beta$) becomes very small and when $\mu$  is
positive, we obtain an additional  reduction of the $hb\bar b$ coupling by a factor
$\approx 1- \Delta_b/(\tan \alpha_{\rm eff} \tan \beta)$. This combination of
parameters leads to a reduction of the decay rate\footnote{Note that in this
small $\alpha_{\rm eff}$ scenario, the rate for the $h\to \tau^+ \tau^-$ 
channel will also be suppressed since $g_{h\tau \tau} \propto
-\sin\alpha_{\rm eff}/\cos\beta$. In turn, there is no significant change
by $\Delta_\tau$ corrections, that are similar to $\Delta_b$ for the electro-weak
part but much smaller (they are now included in the program {\tt HDECAY 5.0}
\cite{Djouadi:1997yw}).}
for $h\to b\bar b$ thereby enhancing all other channels, including $h \to
\gamma\gamma$. This would explain a possible excess in the rate of the $\gamma\gamma$ channel 
without any modification to the $gg\to h$ production rate or the $h\to \gamma \gamma$
branching fraction. In turn, this effect should have no impact on the ratio of
decay widths, $D_{\gamma \gamma}$, which does not depend on the total Higgs width. 
The dependence of $R_{bb}$ on $\mu \tan \beta $ through the $\Delta_b$ correction
and on $\sin \alpha_{\rm eff}$, {\it i.e.} on the $hb\bar b$ coupling without the vertex 
corrections, are shown in Figure~\ref{fig:Rbb}. For small values of 
$\sin \alpha_{\rm eff}$ the variations of $R_{bb}$ from the $\Delta_b$ term are 
enhanced, increasing or decreasing its value depending on the sign of $\mu$.

The total width can also be modified by additional decay channels to SUSY
particles. Because of the LEP2 constraints, the only possible channel for the
$h$ boson is the invisible decay  into pairs of the
lightest  neutralinos $h \to  \chi^0_1  \chi^0_1$. The invisible width can be
important for $M_{\tilde \chi^0_1} <$ 60~GeV and  for not too large $M_1$ and
$|\mu|$ values, and may substantially suppress the decays into SM particles.
This potentially large effect can  be revealed by a combined study of the
individual signal strength values in the various visible Higgs decay channels,
since the changes in these channels are correlated, but not from the ratio
$D_{\gamma \gamma}$. Upper bounds on the invisible decay rate have been obtained 
from the measured signal strengths~\cite{Espinosa:2012vu, Dobrescu:2012td}.

\begin{figure}[h!]
\vspace*{-5mm}
\begin{center}
\begin{tabular}{cc}
\hspace*{-0.60cm} \includegraphics[width=7.5cm]{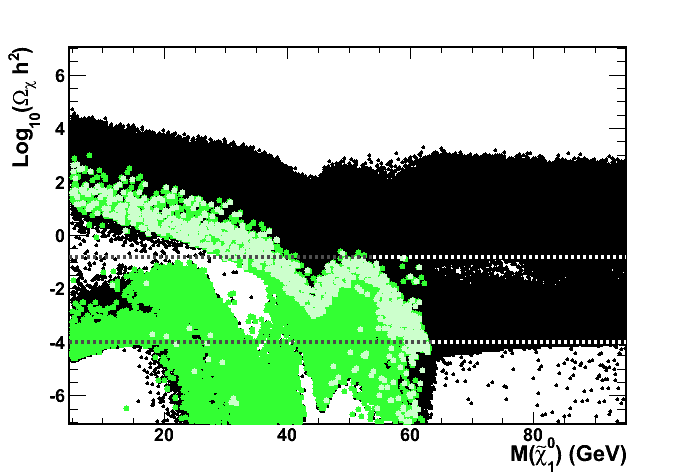} &
\hspace*{-0.90cm} \includegraphics[width=7.5cm]{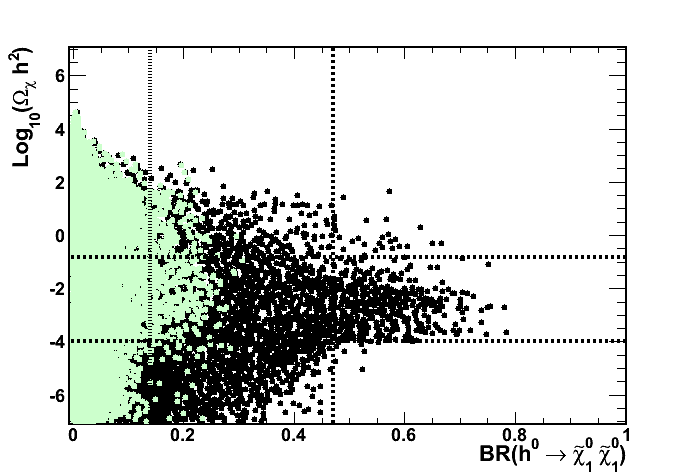} \\
\end{tabular}
\end{center}
\vspace*{-7mm}
\caption{\small The neutralino relic density $\log_{10} (\Omega_{\chi} h^2)$ as a function of 
$M_{\chi_1^0}$ (left) and BR($h \to  \chi^0_1  \chi^0_1$) (right) for the accepted set of pMSSM points 
(black dots), those with BR$(h\to \chi_1^0 \chi_1^0) \geq 15\%$ (green dots) 
and those compatible at 90\% C.L. with the Higgs data (light green dots). The horizontal lines show the 
constraint imposed on $\Omega_{\chi} h^2$ and the vertical lines on the panel on the right the 68\% and 
95\% C.L. constraints on the Higgs invisible decay branching fraction obtained by~\cite{Dobrescu:2012td}.} 
\label{fig:DM}
\vspace*{-1mm}
\end{figure}

The neutralino LSP, with such small mass, would have the relic density required by the 
WMAP results, since it will annihilate efficiently through the exchange of the $h$ boson. 
However, in this case the invisible branching fraction should be small. This is exemplified in
Figure~\ref{fig:DM} where $\log_{10} (\Omega_{\chi} h^2)$ is shown as a function of $M_{\chi_1^0}$
for the accepted set of pMSSM points and for those which have BR$(h\to \chi_1^0 \chi_1^0) \geq 15\%$, 
close to the 68\% C.L. upper limit obtained in~\cite{Dobrescu:2012td}. 
As can be seen only a small area in the region 30 $\lsim M_{\chi_1^0} \lsim$ 60~GeV fulfils this last 
condition and the $\Omega_{\chi} h^2$ constraint. 

We consider now the contributions of SUSY particles to the $\gamma \gamma$
branching fraction and, eventually, to the $gg\to h$ amplitude \cite{Hpp,Htau}. 
Even though the individual contributions give corrections of $\cal{O}$(10~\%)
and in some cases more, it is interesting to observe that different corrections
can sum up, resulting in sizeable overall shifts of the branching fractions
compared to their SM values. These contributions come from light scalar top and
bottom quarks, staus and charginos, as briefly summarised below. 

$a)$ {Stop squark loops}: as already discussed
in~\cite{Arbey:2011ab,Arbey:2012dq},  the Higgs mass constraint requires a very
large SUSY scale $M_S = \sqrt{ m_{\tilde t_1} m_{\tilde t_2}}$ and/or a large
value of the stop  mixing parameter $X_t=A_t -\mu/\tan \beta$ to maximise the
radiative corrections to $M_h$. If $\tilde t_1$ is light, $M_{\tilde t_1} \lsim$
500~GeV, the mixing term must be $X_t \approx \sqrt 6 M_S$ to obtain $M_h
\approx$ 126~GeV. In this case the $h \tilde t_1 \tilde t_1$ coupling, that is
also proportional to $X_t$, becomes large and leads to sizeable stop loop
contributions to the induced Higgs couplings to gluons and photons. However, a
$h\to \gamma\gamma$ rate enhancement is compensated by a suppression of
the  $gg\to h$ production cross section. 

$b)$ {Light sbottom squarks}:  a light right--handed $\tilde b_R$ state,  as
$\tilde b_L$ which belongs to the same iso-doublet as $\tilde t_L$ should be
heavier, does not conflict with the $M_h$ value since the radiative corrections
from the sbottom sector are in general small. For $M_{\tilde b_1} \lsim$
400~GeV, it contributes to the $hgg$ vertex and  slightly enhances the $gg\to h$
production rate. In turn, it would have little impact on the $h\to \gamma\gamma$
rate because of the largely dominating $W$ loop and the small $\tilde b_1$
electric charge. Hence, the $gg\to h \to \gamma\gamma$ rate could be slightly
enhanced by light sbottoms. 

$c)$ {Light $\tilde \tau$ sleptons}: they have received most of
the attention in the literature as it might lead to the largest contributions,
see {\it e.g.} Refs.~\cite{Htau}.  For low
stau mass parameters  $M_{\tilde \tau_L},  M_{\tilde \tau_R}
\approx $ a few 100~GeV, and large stau mixing parameter $X_\tau= A_\tau - \mu
\tan \beta$, with $\tan \beta \approx 60$ and $|\mu|$=500--1000~GeV leading to
$|X_\tau| \approx$ 30--60~TeV, the lighter $\tilde \tau_1$ state has a mass
close to the LEP2 bound, $M_{\tilde \tau_1} \approx$ 100~GeV and its coupling to
the $h$ boson, $g_{h\tilde \tau \tilde \tau} \propto M_\tau X_\tau$, is large.
The $\tilde \tau_1$ contribution, proportional to  $M^2_\tau X^2_\tau/ M_{\tilde
\tau_1}^2 M_{\tilde \tau_2}^2$, can be large enough to significantly increase 
BR($h\to \gamma \gamma$)~\cite{Htau} with a  change of up to 50\%, for extreme
choices of the parameters.

$d)$ {Chargino loops}: the Higgs couplings to charginos are very small if these
are pure winos or higgsinos, and maximal for states with equal higgsino--wino
mixture. Contrary to the scalar case, where the loop contributions are damped by
$1/\tilde M^2$, the chargino contributions to the $h\to \gamma \gamma$ amplitude
are damped only by $1/M_{\tilde \chi_i^\pm}$ factors, so that the decoupling of the
charginos from the  vertex occurs more slowly. However, for a chargino  mass
$M_{\tilde \chi_1^\pm} \gsim$ 100~GeV and maximal couplings to the $h$ boson,
the corrections to the $h\to \gamma \gamma$ rate do not exceed the 10--15\%
level.  The sign of the correction depends on the sign of $\mu$, with the
enhancement occurring for $\mu>0$, as for the $\Delta_b$ correction.

\subsection{Constraints on MSSM parameters} 

\begin{figure}[ht!]
\vspace*{-3mm}
\begin{center}
\begin{tabular}{cc}
\hspace*{-0.60cm} \includegraphics[width=7.5cm]{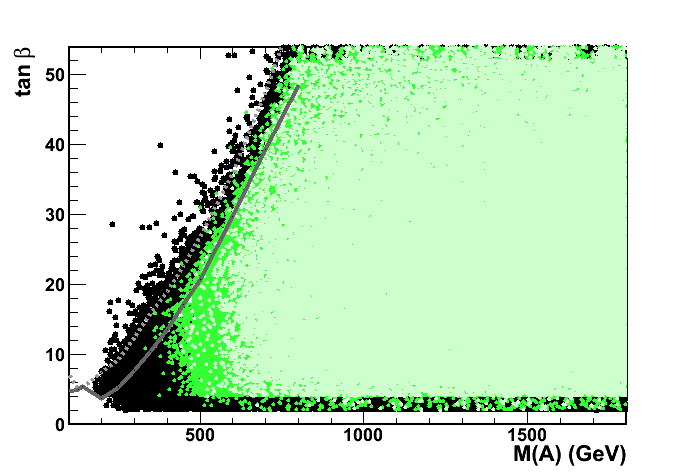} &
\hspace*{-0.90cm} \includegraphics[width=7.5cm]{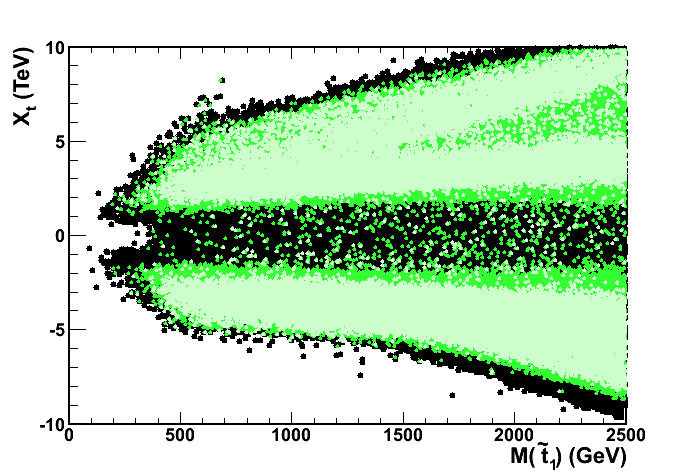}  \\ 
\hspace*{-0.60cm} \includegraphics[width=7.5cm]{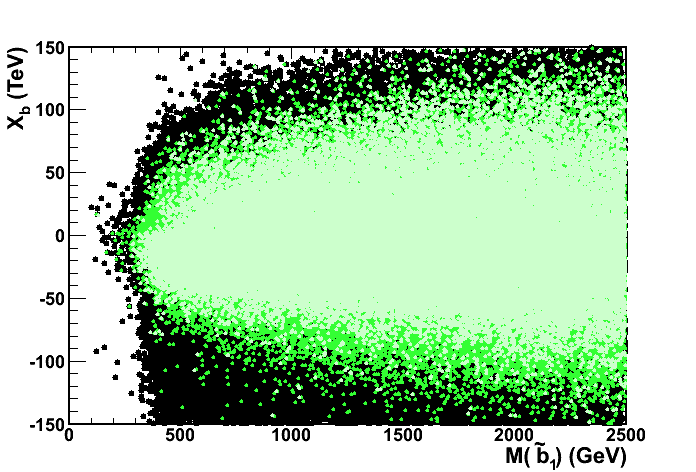} &
\hspace*{-0.90cm} \includegraphics[width=7.5cm]{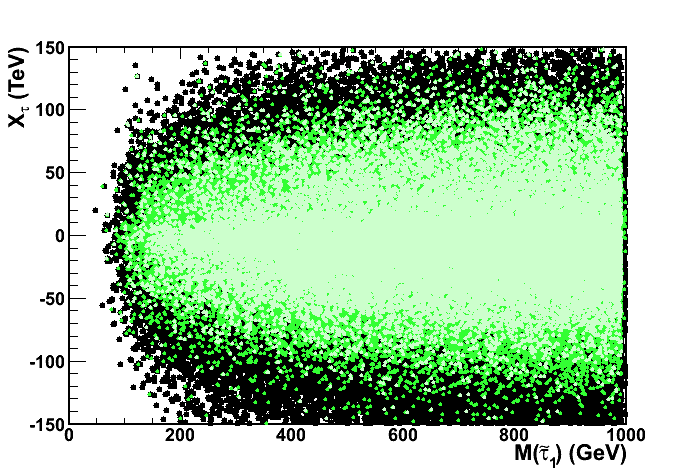} \\
\hspace*{-0.60cm} \includegraphics[width=7.5cm]{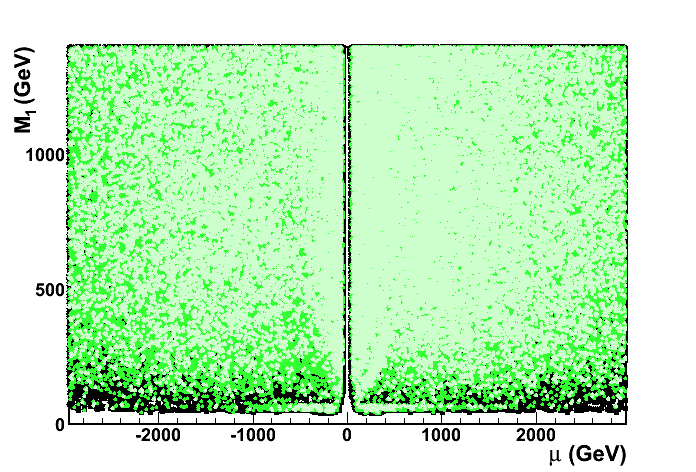} &
\hspace*{-0.90cm} \includegraphics[width=7.5cm]{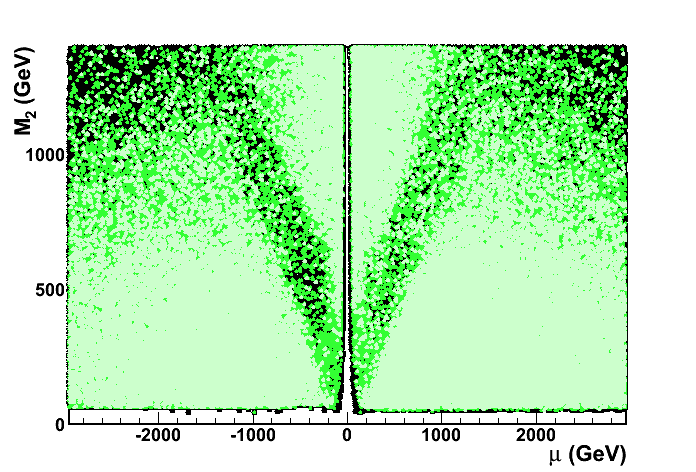} \\
\end{tabular}
\end{center}
\vspace*{-5mm}
\caption{\small Distributions of the pMSSM points in various pMSSM planes.
The black dots show the accepted pMSSM points with $M_h >$ 114~GeV, those 
in dark (light) green the points compatible  with the mass and rate 
constraints of Table~\ref{tab:input} at 90\% (68\%) C.L. On the $[M_A, \tan \beta]$ 
histogram (upper left), the 95\% C.L. expected (dotted line) and observed 
(continuous line) limit from the $H/A \rightarrow \tau \tau$ search of 
Ref.~\cite{CMS-12-050} are superimposed.}
\label{fig:pdf2D}
\end{figure}

In order to study the constraints on the MSSM parameters deriving from the Higgs data, 
we compute the $\chi^2$ probability for the accepted points  with $M_h\! >\! 114$ GeV and 
select those compatible at 90\% and 68\% C.L. with the Higgs constraints of Table~\ref{tab:input}. 
Results are summarised in Figure~\ref{fig:pdf2D}, where we show the accepted pMSSM points
with $M_h >$ 114~GeV  and those compatible with the observed $h$  mass and
signal strengths at the 90\% and 68\% C.L. in the  $[M_A, \tan \beta]$, 
$[M_{\tilde t_1}, X_t]$, $[M_{\tilde b_1}, X_b]$, $[M_{\tilde \tau_1}, X_{\tau}]$,  
$[\mu, M_{1,2}]$ planes.

Given the present statistical accuracy of the LHC results, the 90\% C.L. 
regions, which contain 28\% of the accepted points, have little discriminant
power since all the measured signal  strengths agree with the SM expectations at
this confidence level. On the contrary, the 68\% C.L. regions,  containing
7.7\% of the accepted points, clearly highlight specific regions in the chosen parameter 
sets, where the discrimination is driven mostly by $M_h$ and the 
interplay of the $\mu_{\gamma \gamma}$ value and the  $\mu_{bb}$ and $\mu_{\tau
\tau}$ limits. 

\begin{figure}[ht!]
\vspace*{-3mm}
\begin{center}
\hspace*{-0.50cm} \includegraphics[width=7.5cm]{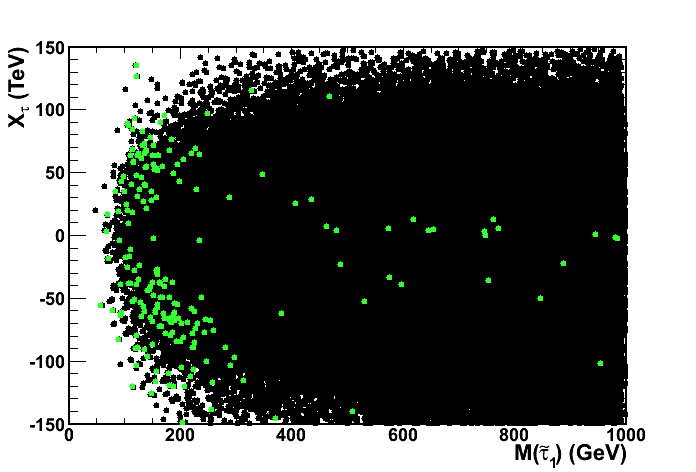}
\end{center}
\vspace*{-5mm}
\caption{\small Distributions of the pMSSM points in the $[M_{\tilde \tau_1}, X_{\tau}]$ plane. 
The black dots show the accepted pMSSM points with $M_h >$ 114~GeV, those in dark green the points 
compatible with the $D_{\gamma \gamma}$ constraint of Table~\ref{tab:input} at 90\% C.L.}
\label{fig:PdfD}
\end{figure}

While the $\mu_{XX}$ values are sensitive to corrections to both the Higgs width
and the loop effects to the  $hXX$ couplings, the $D_{XX}$ values are only
sensitive to the latter, since the width effect gets cancelled  in the ratio.
The current LHC accuracy does not yet provide sensitivity to the  bulk of these
loop effects, expected  to be at the $\cal{O{\mathrm{(10~\%)}}}$ level. Using
the $D_{\gamma \gamma}$  ratio as a constraint at the  90\% C.L, only 0.4\% of
the accepted points are kept. These consists mostly of to points with the 
$\tilde{\tau}_1$ mass in the range between the LEP2 limit and $\sim$200~GeV and 
intermediate values of $X_{\tau}$ (see Figure~\ref{fig:PdfD}), correspond to the scenario 
c) discussed above in Section 2.1, where the $\gamma \gamma$ rate is enhanced by light 
$\tilde{\tau}$ loops. 
Finally, we compare the fraction of accepted
pMSSM points, with $M_h >$ 114~GeV, compatible at 68\% C.L. for the full set of 
observables, which is 7.7\% when we consider the theory systematics and becomes 
0.2\% and 0.1\% for the $M_h$, $\mu_{\gamma \gamma}$, $\mu_{ZZ}$, $\mu_{WW}$ set of 
observables and the full set of observables, without accounting for the production 
cross section uncertainties, respectively.

As can be observed from the $[M_A,\tb]$ plot, the data prefer the decoupling
regime with $M_A \gsim 400$ GeV  for all $\tb$ values and even higher $M_A$ at
large $\tb$. There are however some exceptions and a few points still survive
the strong CMS limit from the $H/A \to \tau^+ \tau^-$ negative search as will be
discussed in more detail shortly. 

As already discussed in Refs.~\cite{Arbey:2011ab,Arbey:2012dq} and elsewhere, 
the lighter stop state can still have a mass of about 500 GeV, but a  strong
stop mixing, $X_t \approx \sqrt 6 M_S$,  is then needed in order to accommodate
the $M_h =126 \pm 3$ GeV value.  Positive values of $X_t \approx A_t$ are
slightly favoured as they allow a better maximisation of the $M_h$ value.  In
the sbottom case, the region with small $M_{\tilde b_1}$ and moderate mixing is
favoured as it leads to light sbottoms that would slightly enhance the $gg\to
h\to \gamma\gamma$ rate. For increasing $M_{\tilde b_1}$ values, the mixing
parameter $X_b \approx -\mu \tb$ tends to be larger  which increases the
$\Delta_b$ corrections and, hence, changes the rate of $R_{bb}$ in the regime
where the $hb \bar b$ coupling is not SM--like as discussed
previously. 

In the $[M_{\tilde \tau_1}, X_\tau]$ plane, a region preferred by the data  is 
the small area with $M_{\tilde \tau_1}=100$--200 GeV, which results in an
enhancement of the $h\to \gamma\gamma$ rate.  At large values of $M_{\tilde
\tau_1}$ for which the stau does not contribute anymore to the $h\gamma\gamma$
vertex, there is still a preference for large $X_\tau$ values but this is mainly
due to the fact that, at large $\tan\beta$, $X_\tau \approx X_b \approx -
\mu \tan\beta$ and, thus, the rate $R_{bb}$ is again affected.

In the $[M_1, \mu]$ and $[M_2, \mu]$  planes, the trend is again mainly driven by
the $\Delta_b$ correction, as the electro-weak SUSY corrections to this
quantity involve several different terms:  a term $\propto A_t \mu \tb$ from
stop contributions, and terms  $\propto M_2 \mu \tb$ and $\propto M_1 \mu \tb$
from the wino and the bino contributions. An exception is  for a very small area
with $|\mu| \approx M_2 \approx 100$ GeV where charginos contribute directly to the
$h\gamma \gamma$ vertex.  

Hence, sbottom mixing plays a  major role in  this analysis as it affects 
strongly the $h \to b\bar b$  decay rate and hence the   branching  fractions
for all other decay channels. This is the reason why the behaviour is rather
different from what was observed with the summer data with $\approx 10$ 
fb$^{-1}$, since, the first result on the signal strength in the $h\to b\bar b$
channel was below unity (and in Ref.~\cite{Arbey:2011ab} the CDF/D0 data were
not included).

\subsubsection{Identification of the observed Higgs state in MSSM}

It has been advocated that the observed 126~GeV particle could indeed be the
heavier $H$  boson~\cite{H=observed}.  This may occur at  low values of $M_A$ 
($\approx 100$--120 GeV), and moderate values of $\tan \beta$ ($\approx 10$). 
In this scenario, the $H$ particle has approximately SM--like properties, while the $h$ boson
has suppressed  couplings to vector bosons and a mass of order 100~GeV or below. In
Ref.~\cite{Arbey:2012dq}, we  performed a dedicated scan for this region of
parameter space and found that only $\approx 2 \times 10^{-5}$ of the generated 
points would remain after imposing the LHC data constraints. These points were
then excluded by applying the constraints from flavour  physics. The most
efficient constraint to this scenario were the $A/H \to \tau^+ \tau^-$  limits
obtained by the ATLAS and CMS collaborations. This search has been updated by
the CMS  collaboration based on 12 fb$^{-1}$ of 8 TeV data and the results exclude 
values of $\tan \beta \gsim 5$ in the entire mass range $90 <M_A< 250$~GeV. 
\begin{figure}[h!]
\begin{center}
\vspace*{-5mm}
\hspace*{-0.80cm} \includegraphics[width=7.5cm]{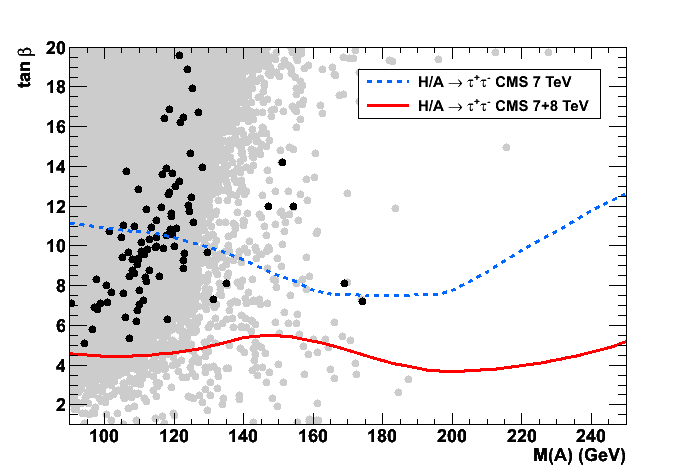}
\end{center}
\vspace*{-5mm}
\caption{\small The parameter space $[M_A, \tan \beta]$ with points for the heavier $H$ boson
to be observed with a mass in the interval 123--129 GeV (light grey points) and after 
flavour and dark matter relic density $10^{-4} <  \Omega h^2 < 0.155$ constraints 
(black points). None of these points have rates compatible with those 
of Table~1. The CMS excluded regions from the 2011 and 2012 $H/A \to \tau^+ 
\tau^-$ searches are shown by the dashed blue and continuous red lines, respectively.}
\label{fig:MA}
\vspace*{-3mm}
\end{figure}

This new result excludes this scenario as is shown in
Figure~\ref{fig:MA} where we zoom in the $[M_A, \tan \beta]$ plane for low values
of the input parameters\footnote{In Ref.~\cite{Arbey:2012dq}, a  few  points allowed 
in this scenario were ruled out by the  $b\to s\gamma$
constraint. In addition, these points did not satisfy the WMAP constraint  of
$10^{-4} <  \Omega h^2 < 0.155$ \cite{Komatsu:2010fb} when  accounting for
theoretical  and cosmological uncertainties \cite{Arbey:2008kv}.}. The small
region in which the $H$ boson was allowed to be the observed state  (green
points) by the previous  $H/A \to \tau^+ \tau^-$ CMS search (dashed blue line), is 
excluded by the new data. In quantitative terms, we observe no point in this scenario 
to comply with the flavour, dark matter and 90\% C.L. for the Higgs data, which corresponds 
to a probability of less than $3 \times 10^{-8}$ for our scan points to realise this scenario 
even before imposing the latest CMS $\tau\tau$ search limits. Conversely, lifting the Higgs 
rate constraints and imposing the  $\tau\tau$ limit leaves us with no viable point for this scenario.

We also note that these new limits also
exclude the so--called ``intense  coupling regime" \cite{intense}, where the
three neutral Higgs bosons could be light and  close in mass (in
Ref.~\cite{Arbey:2012dq}, a very small area of the parameter space at $M_A
\approx$ 140~GeV and $\tan \beta \approx 8$ was still left out). 

\subsubsection{$H/A$ decays into SUSY particles}

There is however a caveat to these $H/A \to \tau^+ \tau^-$ constraints. 
First,  large $\Delta_b$ corrections could significantly enhance the $H/A \to
b\bar b$ decay widths and hence suppress the branching ratio BR($H/A \to \tau^+
\tau^-)$ to make the LHC constraint less efficient. Some (not too large) values
of $\tan \beta$ that are presently excluded could be then resurrected. However,
this can occur only for very large $\Delta_b$ values, ${\cal O}(1)$,  and hence 
extreme choice of the pMSSM parameters. 
\begin{figure}[ht!]
\begin{center}
\vspace*{-3mm}
\begin{tabular}{cc}
\hspace*{-0.60cm} \includegraphics[width=7.5cm]{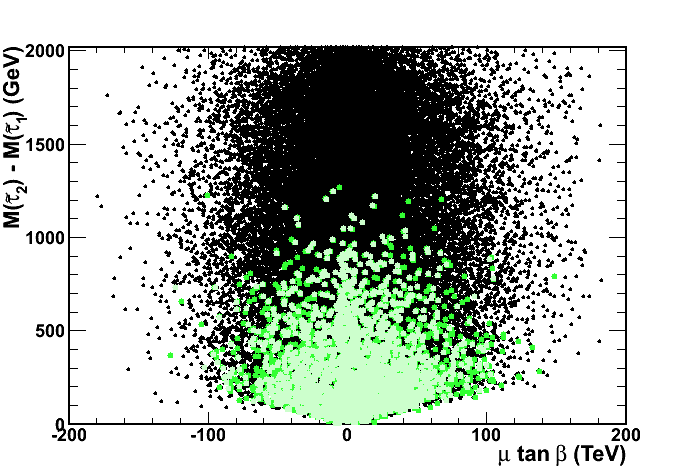} &
\hspace*{-0.90cm} \includegraphics[width=7.5cm]{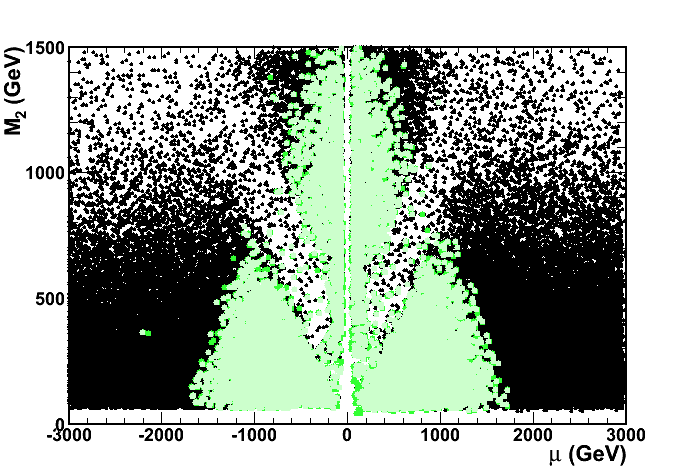} \\
\end{tabular}
\end{center}
\vspace*{-7mm}
\caption{\small Decays of the $H$ boson into SUSY particles. The left panel shows the 
allowed region for the $H \to \tilde \tau_1 \tilde \tau_{1,2}$ decay. 
The black points indicate the area in which the decay is kinematically possible, 
the dark green those with branching ratios larger than 15\% and the light green the subset fulfilling 
also the constraints of Table~\ref{tab:input} at 90\% C.L. The right panel shows the same for the 
$H \to \chi \chi$ where $\chi \chi$ indicates any pair of charginos or neutralinos. The black 
points indicate the accepted pMSSM points, the dark green those with branching ratios larger than 15\% and 
the light green the subset fulfilling also the constraints of Table~\ref{tab:input} at 90\% C.L}
\label{fig:H}
\end{figure}
Another possibility leading to the suppression of the $H/A \to
\tau^+\tau^-$ rate is when the decay channels into SUSY particles are kinematically
accessible. This is  particularly important in the case of the decays $H\to
\tilde \tau_1 \tilde \tau_1$  and $H/A \to \tilde \tau_1 \tilde \tau_2$ (because
of CP invariance the $A$ boson cannot decay into two sfermions of the same
nature) and to a lesser extent $H/A \to \chi_i^0\chi_j^0$ and $\chi_i^+ \chi_j^-$
which can be significant for not too large values of  $\tan \beta$ for which
the total $H/A$ widths are not too strongly enhanced.   
This is shown in Figure~\ref{fig:H} where the points having a branching fraction  BR$(H \to
\tilde \tau_1 \tilde \tau_{1,2})$ larger than 15\% are displayed in the plane 
$[M_{\tilde \tau_2} - M_{\tilde \tau_1},\mu \tan \beta]$. 
This is typically the area in which we have light staus with large
couplings to the Higgs, yielding also an enhancement of $h\to \gamma\gamma$. The branching 
fractions into charginos and  neutralinos are less
significant even for $\tan \beta \lsim 10$.   The invisible decays $H/A \to
\chi_1^0  \chi_1^0$ have more phase space, but the rate is generally small as
the LSP has to be bino--like if is light enough and it thus couples only weakly
to the Higgs bosons (Figure~\ref{fig:H}).  

\subsubsection{Constraints for DM direct detection} 

Finally, we should note that the Higgs data have also an impact on direct dark
matter searches, which are now starting to probe the bulk of the region of the
neutralino-nucleon scattering cross section predicted by the
MSSM. In particular, the latest results reported by the XENON collaboration
improved the earlier 95\% C.L. limit by a factor of $\approx 4$. 

\begin{figure}[ht!]
\vspace*{-4mm}
\begin{center}
\includegraphics[width=7.5cm]{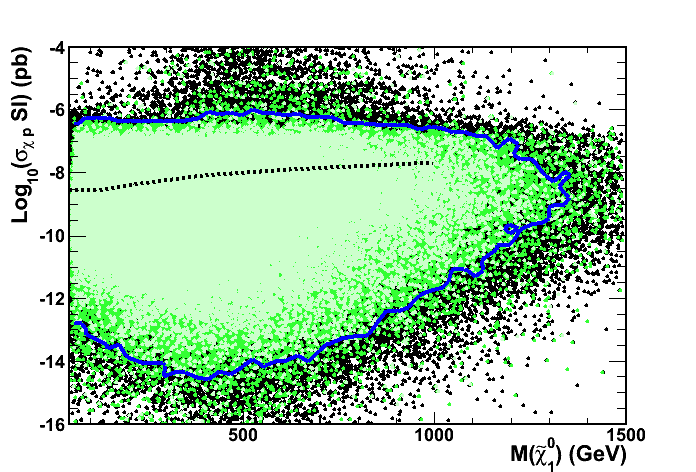} \\
\end{center}
\vspace*{-7mm}
\caption{\small $\chi$-$p$ scattering cross section as a function of the $\chi^0_1$ mass. 
The black dots represent valid pMSSM points, the dark grey dots the subset of points compatible 
at 90\% C.L. with the LHC Higgs results and the light grey dots compatible at 68\% C.L. 
The region enclosed by the grey continuous line contains 99.5\% of the points compatible at 
90\% C.L. with the LHC Higgs results. The dashed line represents the 95\% C.L. upper limit 
contour set by the XENON100 experiment with 225 live days of data.}
\label{fig:DD}
\vspace*{-3mm}
\end{figure}

We compare the new XENON limit with the predicted spin--independent $\chi$--$p$ cross sections as 
a function of the LSP mass for the points fulfilling various selections in Figure~\ref{fig:DD}. 
The XENON limit removes 28\% of the accepted MSSM points before the constraints from the LHC
Higgs results are applied. This fraction decreases to 24\% and 15\% when we
restrict to the points compatible with the measured Higgs mass and rates at,
respectively, the 90\% and 68\% C.L. This indicates that the pMSSM points favoured
by the LHC Higgs results, tend to have a lower $\chi$--$p$ scattering cross section,  
as a result of the large value of $M_A$ that they imply. 

%\newpage
\section{Conclusions}
\label{sec4}

In this paper, we have updated the study of the phenomenological MSSM  performed
in Ref.~\cite{Arbey:2012dq} by including the  new experimental  data recently released. 
We performed a $\chi^2$  probability analysis based on a sample of $2.0
\times 10^8$ generated pMSSM points and presented the regions of the relevant 
MSSM parameters which show agreement at the 68\% and 90\% C.L. with the updated
LHC results. The new, and more precise, ATLAS and CMS  data for the decay
channels $h\to \gamma \gamma$, $W^+W^-$, $ZZ$, $b\bar b$ and $\tau^+ \tau^-$, 
the updated CMS constraints from the $H/A \to \tau^+ \tau^-$ search mode, and the new 
LHCb result for the  $B_s^0 \to \mu^+ \mu^-$ decay branching fraction have a significant
impact on the pMSSM parameter space.

Our earlier results stay qualitatively the same and are even strengthened.  In
particular, we have shown that the possibility of being outside the decoupling
regime for the MSSM Higgs sector by, for instance, allowing the heavier CP--even
$H$ state to be the observed Higgs particle, is now being ruled out. The
scenario in which the total $h$ boson decay width is suppressed, in particular
when the $h b\bar b$ coupling is not SM--like even for $M_A \gg M_Z$, which
enhances the branching fractions for some of the channels still plays a role in view  
of the increased statistical significance of a possible enhancement in the rate of 
the $h \to \gamma \gamma$ decay channel. This also suggests the possibility of some 
light supersymmetric particles, such as staus, charginos and third generation squarks, 
contributing to the $h\gamma\gamma$ loop-induced vertex.
Nevertheless, for not too extreme choices of the pMSSM parameters, we find that
the contributions of the SUSY particles  to the $h\to \gamma\gamma$  branching
fraction should not exceed, in general, the $\approx 20\%$ level. 

\subsection*{Acknowledgements}

We thank N. Berger and M.~Spira for discussion, M.~Schumann and P.~Beltrame for providing 
the XENON results in numerical form. AD thanks the CERN TH unit for hospitality.


\begin{thebibliography}{99}

\bibitem{ATLAS:2012zz}
  G.~Aad {\it et al.}  [ATLAS collaboration],
  %``Observation of a new particle in the search for the Standard Model Higgs boson with the ATLAS detector at the LHC,''
  Phys.\ Lett.\ B716 (2012) 1.
%  [arXiv:1207.7214 [hep-ex]].
  %%CITATION = ARXIV:1207.7214;%%
  
\bibitem{CMS:2012zz}
  S.~Chatrchyan {\it et al.}  [CMS collaboration],
  %``Observation of a new boson at a mass of 125 GeV with the CMS experiment at the LHC,''
  Phys.\ Lett.\ B716 (2012) 30.
%  [arXiv:1207.7235 [hep-ex]].
  %%CITATION = ARXIV:1207.7235;%%

\bibitem{ATLAS-2012-168}
  [ATLAS Collaboration], Note ATLAS-CONF-2012-168.

\bibitem{ATLAS-2012-169}
  [ATLAS Collaboration], Note ATLAS-CONF-2012-169.

\bibitem{CMS-12-015}
  [CMS Collaboration], Note CMS PAS HIG-2012-015.

\bibitem{CMS-12-016}
  [CMS Collaboration], Note CMS PAS HIG-2012-016.

\bibitem{ATLAS-2012-158}
  [ATLAS Collaboration], Note ATLAS-CONF-2012-158.

\bibitem{ATLAS-2012-161}
  [ATLAS Collaboration], Note ATLAS-CONF-2012-161.

\bibitem{CMS-12-044}
  [CMS Collaboration], Note CMS PAS HIG-2012-044.


\bibitem{ATLAS-2012-160}
  [ATLAS Collaboration], Note ATLAS-CONF-2012-160.

\bibitem{CMS-12-043}
  [CMS Collaboration], Note CMS PAS HIG-2012-043.


\bibitem{CMS-12-050}
  [CMS Collaboration], Note CMS PAS HIG-2012-050.


\bibitem{Aaij:2012hcp}
  R.~Aaij {\it et al.}  [LHCb Collaboration],
  CERN-PH-EP-2012-335.

\bibitem{Aprile:2012nq}
  E.~Aprile {\it et al.}  [XENON100 Collaboration],
  Phys.\ Rev.\ Lett.\ 109 (2012) 181301.
%  arXiv:1207.5988 [astro-ph.CO].

%%%%%%%%%%%%%%%%%%%%%%%%%%%%%%%%%%%%%%%%%%%%%%%%%%%%%%%%%%%%%%%%%%%%%

\bibitem{Arbey:2012dq}
  A.~Arbey, M.~Battaglia, A.~Djouadi and F.~Mahmoudi,
  %``The Higgs sector of the phenomenological MSSM in the light of the Higgs boson discovery,''
  JHEP 1209 (2012) 107.%  arXiv:1207.1348 [hep-ph].
  %%CITATION = ARXIV:1207.1348;%%


\bibitem{Arbey:2011ab}
  A.~Arbey et al., %, M.~Battaglia, A.~Djouadi, F.~Mahmoudi and J.~Quevillon,
  %``Implications of a 125 GeV Higgs for supersymmetric models,''
  Phys.\  Lett.\  B708 (2012) 162.
%  [arXiv:1112.3028 [hep-ph]].
  %%CITATION = ARXIV:1112.3028;%%

\bibitem{tools} 
  A. Arbey, M. Battaglia and F. Mahmoudi,  Eur.\ Phys.\ J.\ C72 (2012) 1847; 
  {\it Eadem}, Eur.\ Phys.\ J.\ C72 (2012) 1906.

\bibitem{Tevatron:2012zz}
  The CDF and D0 Collaborations,
  %``Updated Combination of CDF and D0 Searches for Standard Model Higgs Boson Production with up to 10.0 fb-1 of Data,''
  FERMILAB-CONF-12-318-E.
  %%CITATION = FERMILAB-CONF-12-318-E;%%

\bibitem{Mahmoudi:2012un}
  F.~Mahmoudi, S.~Neshatpour and J.~Orloff,
  %``Supersymmetric constraints from $B_s -> \mu^+\mu^-$ and $B -> K* \mu^+\mu^-$ observables,''
  JHEP 1208 (2012) 092.
%  [arXiv:1205.1845 [hep-ph]].

\bibitem{bsgamma} 
  Y. Amhis {\it et al.,} (HFAG collaboration),  arXiv:1207.1158[hep-ex]. 

\bibitem{Djouadi:1997yw}
  A.~Djouadi, J.~Kalinowski and M.~Spira,
  Comput.\ Phys.\ Commun.\  108 (1998) 56.

\bibitem{THU}  
  S. Dittmaier {\it et al.}, [LHC Higgs cross section working group],  
  arXiv:1101.0593 [hep-ph]; J.~Baglio and A. Djouadi, JHEP 1103 (2011) 055. 

\bibitem{Djouadi:2012he}
  A.~Djouadi,
  %``Precision Higgs coupling measurements at the LHC through ratios of production cross sections,''
  arXiv:1208.3436 [hep-ph].
  %%CITATION = ARXIV:1208.3436;%%

\bibitem{vanishing} 
  M. Carena, S. Heinemeyer, C. Wagner and G. Weiglein, 
  Eur.\ Phys.\ J.\ C26 (2003) 601.

\bibitem{Espinosa:2012vu}
  J.~R.~Espinosa, M.~Muhlleitner, C.~Grojean and M.~Trott,
  %``Probing for Invisible Higgs Decays with Global Fits,''
  JHEP 1209 (2012) 126
%  [arXiv:1205.6790 [hep-ph]].
  %%CITATION = ARXIV:1205.6790;%%

\bibitem{Dobrescu:2012td}
  B.~A.~Dobrescu and J.~D.~Lykken,
  %``Coupling spans of the Higgs-like boson,''
  arXiv:1210.3342 [hep-ph].
  %%CITATION = ARXIV:1210.3342;%%

\bibitem{Hpp} A.~Djouadi, V.~Driesen, W.~Hollik and J.~I.~Illana, Eur.\ Phys.\
  J.\ C1 (1998) 149; A.~Djouadi, Phys.\ Lett.\ B435 (1998) 101; 
  A. Djouadi, Phys.\ Rept.\  459 (2008) 1. 

\bibitem{Htau} See e.g., 
  M.~Carena et al.,  JHEP 1207 (2012) 175;
  M.~Carena, I.~Low and C.~E.~M.~Wagner,
  JHEP 1208 (2012) 060;
  G.~F.~Giudice, P.~Paradisi, and A.~Strumia,
  JHEP 1210 (2012) 186;
  U. Haisch and F. Mahmoudi, arXiv:1210.7806 [hep-ph].
  
\bibitem{H=observed} See for instance, S. Heinemeyer, O. Stal and G. Weiglein,
  Phys.\ Lett.\ B710 (2012) 201; P.~Bechtle et al., arXiv:1211.1955 [hep-ph];
  M. Drees,  arXiv:1210.6507 [hep-ph].  

\bibitem{Komatsu:2010fb}
  E.~Komatsu {\it et al.}  [WMAP Collaboration],
  Astrophys.\ J.\ Suppl.\  192 (2011) 18.

\bibitem{Arbey:2008kv}
  A.~Arbey and F.~Mahmoudi,
  Phys.\ Lett.\ B669 (2008) 46;
%  A.~Arbey and F.~Mahmoudi,
  {\it Eadem}, JHEP 1005 (2010) 051.

\bibitem{intense} 
   E. Boos et al., Phys. Rev. D66 (2002) 055004; 
   {\it Eadem}, Phys.\ Lett.\ B578 (2004) 384.

\end{thebibliography}
\end{document}